\documentclass[12pt]{article}
\usepackage{amssymb,amsmath}
\usepackage[noblocks]{authblk}
\usepackage{cite}
\usepackage[top=0.75in, bottom=0.75in, left=0.75in, right=0.75in, dvips]{geometry}
\usepackage{caption}
\usepackage{multicol}

\begin{document} 
\title{The Effective Potential In Non-Conformal\\ Gauge Theories}
\author[1]{F. T. Brandt\footnote{fbrandt@usp.br}}
\author[2]{F. A. Chishtie\footnote{fchishtie@uwo.ca}}
\author[3,4]{D. G. C. McKeon\footnote{dgmckeo2@uwo.ca}}
%
\affil[1]{Instituto de Fisica, Universidade de S$\mathrm{\tilde{a}}$o Paulo, S$\mathrm{\tilde{a}}$o Paulo, SP 05508-900, Brazil}
\affil[2]{Theoretical Research Institute, Pakistan Academy of Sciences (TRIPAS), Islamabad 44000, Pakistan}
\affil[3]{Department of Applied Mathematics, University of Western Ontario, London Canada N6A 5B7}
\affil[4]{Department of Mathematics and Computer Science, Algoma University, Sault St. Marie Canada P6A 2G4, Canada}
\date{\today}
\maketitle
        
\maketitle
\noindent
PACS No.: 14.80Bn\\
KEY WORDS: effective potential, massive scalars

\begin{abstract}
By using the renormalization group (RG) equation it has proved possible to sum logarithmic corrections to quantities that arise due to quantum effects in field theories.  In particular, the effective potential $V$ in the Standard Model in the limit that there are no massive parameters in the classical action (the ``conformal limit'') has been subject to this analysis, as has the effective potential in a scalar theory with a quartic self coupling and in massless scalar electrodynamics.  Having multiple coupling constants and/or mass parameters in the initial action complicates this analysis, as then several mass scales arise.  We show how to address this problem by considering the effective potential in a Yukawa model when the scalar field has a tree-level mass term.  In addition to summing logarithmic corrections by using the RG equation, we also consider the consequences of the condition $V^\prime(v) = 0$ where $v$ is the vacuum expectation value of the scalar.  If $V$ is expanded in powers of logarithms that arise, then it proves possible to show that either $v$ is zero or that $V$ is independent of the scalar.  (That is, either there is no spontaneous symmetry breaking or the vacuum expectation value is not determined by minimizing $V$ as $V$ is ``flat''.)
\end{abstract}

\section{Introduction}

The effective potential \cite{r1,r2,r3,r4} has long been an integral part of both electroweak theory and cosmology. In particular, it has provided a way to endowing non-Abelian vector Bosons with a mass without destroying renormalizability (see any current text; eg Ref \cite{r5})\footnote{We note that an Abelian vector can be given a mass ``by hand''using the Stueckelberg formalism; this includes the $U(1)$ sector of the Standard Model \cite{r6,r7}.}. The possibility that the spontaneous symmetry breakdown responsible for this is entirely due to quantum effects and that all mass scales are a consequence of there being a radiatively induced mass has been introduced in \cite{r1} and further considered in \cite{r8,r9,r10,r11}. 

In these latter two references, the RG equation has been used to sum logarithmic corrections due to radiative effects; the one-loop RG functions plus the classical potential yields the sum of all leading-log (LL) effects, the two-loop RG functions plus the  log independent part of the one-loop effective potential yield the next-to-leading-log (NLL) effects, etc. This approach has been used in the calculation of cross sections \cite{r11}, in relating the bare and renormalized coupling in dimensional regularization \cite{r12}, in thermal field theory \cite{r13} as well as with the effective potential \cite{r14,r15,r16,r8,r9,r10,r11}. 

Summing these radiative effects is made more complicated if in addition to the radiatively induced mass $\mu^2$ there is a tree level mass parameter $m^2$ associated with the scalar field. Since the classical conformal symmetry present if $m^2 = 0$ is broken by quantum effects (at least at the perturbative level) there is no reason for setting $m^2=0$ other than desire for simplicity, though it also is relevant in aspects of the scale hierarchy \cite{r17} and fine-tuning \cite{r18} problems. We are thus motivated to consider the possibility of summing radiative effects when $m^2\neq 0$; we will also consider having multiple couplings.

There have been several approaches to employing the RG to treat situations in which there are multiple mass scales. One of these is to have a number of different radiatively induced mass scales and to have an RG equation associated with each of them \cite{r24,r25,r31}. Another is to retain a single radiatively induced mass scale $\mu^2$ and the rewrite $\log\frac{m_2^2}{\mu^2}$  as $\log\frac{m_2^2}{m_1^2}+\log\frac{m_1^2}{\mu^2}$ to convert the mass scale   $\frac{m_2^2}{\mu^2}$ to $\frac{m_1^2}{\mu^2}$ \cite{r15,r16}. We will employ the later approach.

In addition to arranging perturbative contributions to the effective potential in such a way that the RG equation can be used to sum the LL, NLL etc. contributions, one can also simply write the effective potential as a sum of a log-independent contribution, a contribution linear in the logarithm, a term quadratic in the logarithm etc. Once this is done then a summation can be performed in which not only is the RG equation used, but also the condition that $V$ be minimized at the vacuum expectation value of the scalar field $\phi$. By using these two conditions, it not only proves possible to express the effective potential in terms of its log-independent contribution, but also to show that all dependence of $V$ on the scalar field cancels provided there is spontaneous symmetry breaking. Having the effective potential independent of the scalar field does not exclude the possibility of spontaneous symmetry breaking, it just means that the vacuum expectation value of the scalar does not occur at a local minimum of the effective potential as it is now flat. This flatness does not necessarily mean that the theory is non-interacting (i.e., ``trivial'').  The full effective action for $\phi$ may involve interactions which depend on the gradient of this scalar or interactions between the scalar and other fields. This has been demonstrated in a simple scalar theory with a quartic self coupling, in massless scalar electrodynamics. as well as a massive scalar theory with a quartic self coupling \cite{r19,r20,r21}. Below we will first review the simple massless scalar model and then we will demonstrate that this result also holds in models in which the scalar field has a tree-level mass and there are multiple couplings, first by examining a Yukawa model with a massive scalar and a massless spinor and then the Standard Model.

As a check on the validity of our expression for $V$ we will demonstrate that satisfies the RG equation.

\section{The Effective Potential}

We shall start by reviewing how the RG can be used to sum certain higher loop contributions to the effective potential in a massless scalar theory with a quartic scalar coupling. Starting from the action \cite{r19}
\begin{eqnarray}\label{e1}
S = \int d^4 x \left[\frac 1 2 \left( \partial_\mu \phi \right)^2 - \frac{\lambda}{4!} \phi^4\right]
\end{eqnarray}
we compute the effective potential $V$, which has the form
\begin{eqnarray}\label{e2}
V(\lambda,\phi,\mu)=\sum_{n=0}^{\infty} \sum_{m=0}^{\infty} \lambda^{m+n+1} T_{m+n,m} L^m \phi^4 \equiv
\sum_{n=0}^{\infty} \lambda^{n+1} S_n(\lambda L) \phi^4
\end{eqnarray}
where, if we use the CW renormalization condition \cite{r1}
\begin{eqnarray}\label{e3}
\lambda = \frac{1}{4!} \frac{d^4 V(\phi)}{d \phi^4}\quad \mathrm{when}\;\;\phi = \mu
\end{eqnarray}
then
\begin{eqnarray}\label{e4}
L = \log(\phi^2/\mu^2 ).
\end{eqnarray}
Since changes in the (unphysical) renormalization scale $\mu$ must be compensated by corresponding changes in the other parameters that characterize the theory (in this case $\lambda$ and $\phi$) we find that 
\begin{eqnarray}\label{e5}
\mu \frac{d}{d\mu} V= \left( \mu \frac{\partial}{\partial\mu} 
+ \beta(\lambda) \frac{\partial}{\partial\lambda} + \gamma(\lambda) \phi \frac{\partial}{\partial\phi}  \right) V(\lambda,\phi,\mu) = 0,
\end{eqnarray}
where
\begin{subequations}\label{e6}
\begin{eqnarray}\label{e6a}
\beta = \mu \frac{\partial \lambda}{\partial \mu} = \sum_{k=2}^{\infty} b_k \lambda^k 
\end{eqnarray}
and
\begin{eqnarray}\label{e6b}
\gamma = \mu \frac{\partial \log\phi}{\partial \mu} = \sum_{k=1}^{\infty} g_k \lambda^k. 
\end{eqnarray}
\end{subequations}

Together, Eqs. (\ref{e2},\ref{e5},\ref{e6}) result in the recursive equations
\begin{subequations}\label{e7}
\begin{eqnarray}\label{e7a}
\left[\left(-2+b_2 \xi\right)\frac{d}{d\xi} +\left(b_2+4 g_1\right)\right] S_0(\xi)=0
\end{eqnarray}
\begin{eqnarray}\label{e7b}
&\left[\left(-2+b_2 \xi\right)\frac{d}{d\xi} + (n+1)b_2 +4 g_1 \right] S_n(\xi)  \nonumber \\
&+\displaystyle{
\sum_{m=0}^{n-1} \left[\left(2 g_{n-m} + b_{n-m+2} \xi\right)\frac{d}{d\xi} + (m+1) b_{n+2-m} + 4 g_{n+1-m}\right] S_m(\xi) }=0
\end{eqnarray}
with the boundary condition
\begin{eqnarray}
S_n(0)=T_{n,0} .
\end{eqnarray}
\end{subequations}
The solution $S_n(\xi)$ gives the ${\rm N}^n {\rm LL}$ contribution to $V$ \cite{r14,r8}.

We note that the solution to Eq. \eqref{e5} when only $b_2 \dots b_m,\, g_1\dots g_{m-1}$ are non-zero is not given by having only $S_0\dots S_{m-2}$ contribute to Eq. \eqref{e2}; these RG function coefficients also contribute to all $S_k$ ($k> m-2$). In fact, we see that since $S_n$ has the form
\begin{eqnarray}\label{e8}
S_n = \frac{1}{b_2} \sum_{i=1}^{n+1} \sum_{j=0}^{i-1} \sigma_{i,j}^n \frac{\Lambda^j}{w^i}
\end{eqnarray}
($w=1-b_2\xi/2,\Lambda = \log|w|$) then recursive relations can be found from Eq. \eqref{e7b} for the coefficients $\sigma_{i,j}^n$ \cite{r22}; for example we find that
\begin{eqnarray}\label{e9}
\sigma_{n+1,n}^{n} = \left(\frac{-b_3}{b_2}\right)^n \sigma_{1,0}^{0} 
\end{eqnarray}
with $\sigma_{1,0}^0=1/4!$. By Eq. \eqref{e8}, Eq. \eqref{e9} gives a portion of all $S_n$ ($n>1$) even though with $b_2,\,b_3,\,g_1,\,g_2$ only $S_0$ and $S_1$ can be determined exactly.

We can also examine the general form of the solution to Eq. \eqref{e5} using the method of characteristics (moc) as the RG equation is a linear first-order partial differential equation \cite{r23}. We introduce characteristic functions 
$\bar\mu(s)$, $\bar\lambda(s)$, $\bar\phi(s)$ and $\bar V(s)$ satisfying
\begin{equation}
\frac{d\bar{\mu}(s)}{ds} = \bar{\mu}(s),\quad 
\frac{d\bar{\lambda}(s)}{ds} = \beta(\bar{\lambda}(s)), \quad
\frac{d\bar{\phi}(s)}{ds} = \bar{\phi}(s) \gamma (\bar{\lambda}(s)),\quad
\frac{d\bar{V}(s)}{ds} = 0  \addtocounter{equation}{1} \tag{\theequation a--d} \label{e10}  
\end{equation}
%
with $\bar\mu(0) = \mu$ etc. Upon solving for $\bar\mu(s)$, $\bar\lambda(s)$ and $\bar\phi(s)$ it follows that
\begin{eqnarray}\label{e11}
V(\lambda,\phi,\mu)=V(\bar\lambda(0),\bar\phi(0),\bar\mu(0))
\end{eqnarray}
is a solution to Eq. \eqref{e5}. From Eqs. \eqref{e10} 
we find that
\begin{subequations}\label{e12}
\begin{eqnarray}\label{e12a}
\log\bar\mu(s) = s + c_1(t) 
\end{eqnarray}
\begin{eqnarray}\label{e12b}
\int_{\lambda_0}^{\bar\lambda(s)}\frac{dx}{\beta(x)}=s + c_2(t)
\end{eqnarray}
\begin{eqnarray}\label{e12c}
\log\bar\phi(s)=\int_{\lambda_0}^{\bar\lambda(s)}dx\frac{\gamma(x)}{\beta(x)}+c_3(t),
\end{eqnarray}
\end{subequations}
where $t$ is a parameter associated with the constants of integration arising in Eqs. \eqref{e10} 
\cite{r23}. Using Eqs. \eqref{e12}, we find that since 
\begin{subequations}\label{e13}
\begin{eqnarray}\label{e13a}
c_1(t) - c_2(t) = \log\bar\mu(s) - \int_{\lambda_0}^{\bar\lambda(s)} \frac{dx}{\beta(x)},
\end{eqnarray}
\begin{eqnarray}\label{e13b}
c_3(t)=\log\bar\phi(s) - \int_{\lambda_0}^{\bar\lambda(s)} dx \frac{\gamma(x)}{\beta(x)} 
\end{eqnarray}
\end{subequations}
are independent of $s$, a general solution for $V$ by \eqref{e11} is given by
\begin{eqnarray}\label{e14}
V(\lambda,\phi,\mu)&=&F\left(
\log\mu - \int_{\lambda_0}^{\lambda} \frac{dx}{\beta(x)} , \log\phi - \int_{\lambda_0}^{\lambda} dx \frac{\gamma(x)}{\beta(x)}
\right) \nonumber \\
&\equiv& F(A,B)
\end{eqnarray}
where $F$ is determined by boundary conditions in Eq. \eqref{e5}.

If now we write Eq. \eqref{e2} as
\begin{eqnarray}\label{e15}
V=\sum_{n=0}^{\infty} A_n(\lambda) l^n \phi^4 \;\;\;  (l=\log\phi/\mu)
\end{eqnarray}
where
\begin{eqnarray}\label{e16}
A_n(\lambda)=\sum_{m=n}^{\infty} \lambda^{m+1} 2^{-n} T_{m,n}
\end{eqnarray}
then together Eqs. (\ref{e5},\ref{e15}) lead to
\begin{eqnarray}\label{e17}
\hat A_{n+1}(\eta)=\frac{1}{n+1}\frac{d}{d\eta} \hat A_n(\eta) = \frac{1}{(n+1)!}\frac{d^{n+1}}{d\eta^{n+1}} \hat A_0(\eta)
\end{eqnarray}
where
\begin{subequations}\label{e18}
\begin{eqnarray}\label{e18a}
\eta(\lambda)=\int_{\lambda_0}^{\lambda} dx \frac{1-\gamma(x)}{\beta(x)}
\end{eqnarray}
and
\begin{eqnarray}\label{e18b}
\hat A(\lambda) = A_n(\lambda) \exp\left(4 \int_{\lambda_0}^{\lambda} dx \frac{\gamma(x)}{\beta(x)}\right).
\end{eqnarray}
\end{subequations}
Since
$\sum_{n=0}^{\infty} \frac{a^n}{n!} \frac{d^n}{dx^n} f(x) = f(x+a)$,
Eqs. (\ref{e15},\ref{e17}) together yield \cite{r19}
\begin{eqnarray}\label{e19}
V = \exp\left(4 \int_{\lambda}^{\lambda(\eta+l)} dx \frac{\gamma(x)}{\beta(x)}\right) A_0(\lambda(\eta+l)) \phi^4.
\end{eqnarray}
We note that
\begin{subequations}\label{e20}
\begin{eqnarray}\label{e20a}
\eta+l&=&\int_{\lambda_0}^{\lambda} \frac{dx}{\beta(x)} - \int_{\lambda_0}^{\lambda} dx \frac{\gamma(x)}{\beta(x)}
+\log\phi -\log\mu \nonumber \\
&=& -A+B
\end{eqnarray}
and
\begin{eqnarray}\label{e20b}
\exp\left(4 \int_{\lambda}^{\lambda(\eta+l)} dx \frac{\gamma(x)}{\beta(x)}\right) \phi^4
=\exp\left(4 \int_{\lambda_0}^{\lambda(-A+B)} dx \frac{\gamma(x)}{\beta(x)} +B\right) .
\end{eqnarray}
\end{subequations}
By Eq. \eqref{e20}, $V$ in Eq. \eqref{e19} is of the form Eq. \eqref{e14} and thus the RG equation of Eq. \eqref{e5} is satisfied.

In order to find the function $A_0(\lambda)$ we impose a second condition on $V$ in addition to the RG equation \eqref{e5}. This second condition is
\begin{eqnarray}\label{e21}
\frac{dV(\phi=v)}{d\phi}=0
\end{eqnarray}
where $v$ is the vacuum expectation value (vev) of $\phi$. Substitution of Eq. \eqref{e15} into \eqref{e21} leads to
\begin{eqnarray}\label{e22}
\sum_{n=0}^{\infty} \left((n+1) A_{n+1}(\lambda) + 4 A_n(\lambda)\right) \log^n\left(\frac{v}{\mu}\right) v^3 = 0.
\end{eqnarray}
One solution to Eq. \eqref{e22} is
\begin{eqnarray}\label{e23}
v=0
\end{eqnarray}
in which case there is no spontaneous symmetry breaking. If $v\neq 0$, then we can chose $\mu = v$ leading to
\begin{eqnarray}\label{e24}
A_1(\lambda(v)) = -4 A_0(\lambda(v)).
\end{eqnarray}
The actual value of $\lambda(v)$ is unspecified, and hence Eq. \eqref{e24} is a general functional relation between $A_1(\lambda)$ and $A_0(\lambda)$. When $n=0$ in Eq. \eqref{e17} (which follows from the RG equation)
\begin{eqnarray}\label{e25}
A_1(\lambda) = \frac{\beta(\lambda)}{1-\gamma(\lambda)}\left(\frac{d}{d\lambda}+ \frac{4\gamma(\lambda)}{\beta(\lambda)}\right) A_0(\lambda).  
\end{eqnarray}
Together, Eqs. (\ref{e24},\ref{e25}) lead to
\begin{eqnarray}\label{e26}
\left(\beta(\lambda)\frac{d}{d\lambda}+4\right)A_0(\lambda)=0
\end{eqnarray}
whose solution is
\begin{eqnarray}\label{e27}
A_0(\lambda)=A_0(\lambda_0) \exp\left(-4\int_{\lambda_0}^\lambda \frac{dx}{\beta(x)}\right).
\end{eqnarray}
From Eqs. \eqref{e27} and \eqref{e19} we find that
\begin{eqnarray}
V&=&\exp\left(4\int_{\lambda}^{\lambda(\eta+l)} dx \frac{\gamma(x)}{\beta(x)}\right) A_0(\lambda_0)
\exp\left(-4\int_{\lambda_0}^{\lambda(\eta+l) }\frac{dx}{\beta(x)}\right) \phi^4 \label{e28} \\
&=& A_0(\lambda_0) \exp\left(-4\int_{\lambda_0}^{\lambda(\eta) } \frac{dx}{\beta(x)}\right)
\exp \left(4\int_{\lambda(\eta)}^{\lambda(\eta+l)} dx \frac{\gamma(x) -1 }{\beta(x)}\right) \phi^4. \label{e29}
\end{eqnarray}
But by Eq. \eqref{e18a},
\begin{eqnarray}\label{e30}
\int_{\lambda(\eta)}^{\lambda(\eta+l)} dx \frac{\gamma(x) - 1}{\beta(x)} = -(\eta+l) + \eta = -\log\left(\frac{\phi}{\mu}\right) .
\end{eqnarray}
So that upon substitution of Eq. \eqref{e30} into \eqref{e29}
\begin{eqnarray}\label{e31}
V(\lambda,\phi,\mu)=A_0(\lambda_0) \exp\left(-4\int_{\lambda_0}^{\lambda} \frac{dx}{\beta(x)}\right) \mu^4
\end{eqnarray}
and all dependence of $V$ on $\phi$ drops out.

If instead of combining Eqs. \eqref{e24} and \eqref{e25} to obtain Eq. \eqref{e31}, we had taken Eq. \eqref{e22} to hold for each of the coefficients of $\log^n\left(\frac v \mu \right)$, then we would have
\begin{eqnarray}\label{e32}
A_{n+1}(\lambda) = \frac{-4}{n+1} A_n(\lambda) = \frac{(-4)^{n+1}}{(n+1)!} A_0(\lambda)
\end{eqnarray}
and not just Eq. \eqref{e24}. Now combining Eqs \eqref{e15} and \eqref{e32} we obtain
\begin{eqnarray}\label{e33}
V &=& \sum_{n=0}^{\infty} A_0(\lambda) \frac{(-4)^n}{n!} \phi^4 \nonumber \\
&=& A_0(\lambda) \exp\left(-4 \log\frac \phi \mu\right) \phi^4 \nonumber \\
&=& A_0(\lambda) \mu^4.
\end{eqnarray}
For Eq. \eqref{e33} to satisfy the RG equation \eqref{e5}, $A_0(\lambda)$ is given by Eq. \eqref{e27}.

Having a ``flat'' effective potential in which $V$ is independent of $\phi$ (Eq. \eqref{e31}) does not mean there cannot be spontaneous symmetry breaking; it just means that the vev of $\phi$ cannot be fixed by finding a local minimum of $V$. Having $V$ independent of $\phi$ is not the same as saying that Eq. \eqref{e21} is satisfied at the particular value $\phi=v$. 

We note that a general solution to the RG equation of Eq. (5) is simply
\begin{equation}
V = V_0 \mu^\alpha \phi^K \exp \left[ - \int_{\lambda_{0}}^\lambda \frac{K\gamma(x) + \alpha}{\beta(x)} dx\right] \nonumber
\end{equation}
for arbitrary $K$ and $\alpha$; summing over $K$ and $\alpha$ leads to
\begin{equation}
= V_0 \frac{\mu\phi}{\left(\mu - \exp \left[ -\int_{\lambda_{0}}^\lambda \frac{1}{\beta(x)}dx \right]\right) \left(\phi - \exp \left[ -\int_{\lambda_{0}}^\lambda \frac{\gamma(x)}{\beta(x)}dx \right]\right) } .\nonumber
\end{equation}
Eq. (31) corresponds to taking $K = 0$, $\alpha = 4$ and $V_0 = A_0(\lambda_0)$. However, the general solution for $V$ with arbitrary $K$ and $\alpha$ is not acceptable as it does not have a lower bound, it is not consistent with the initial ansatz of Eqs. (2) and (15), and does not meet the condition of Eq. (21).

After reviewing the situation in which there is a single massless self interacting scalar field, we now examine a model in which the real scalar $\phi$ has a quartic self coupling $\lambda$, a mass m and a Yukawa coupling $g$ with a massless spinor. In this case, the one-loop effective potential is \cite{r26}
\begin{eqnarray}\label{e34}
V^{(1)}=\frac{\left(m^2+\lambda\phi^2/2\right)^2}{4(4\pi)^2}\left(\log\frac{m^2+\lambda\phi^2/2}{\mu^2} -\frac 3 2\right)
-\frac{g^4 \phi^4}{(4\pi)^2}\left(\log\frac{g^2\phi^2}{\mu^2} -\frac 3 2\right),
\end{eqnarray}
when using dimensional regularization. In general $V$ is given by
\begin{eqnarray}\label{e35}
V=\phi^4\sum_{N=L+1}^{\infty}  \sum_{n=0}^{N}  \sum_{L=0}^{\infty}  \sum_{l=0}^{L}
p_{L,l}^{N,n}(r,z) \lambda^{N-n} g^{2n} \Lambda_1^{L-l} \Lambda_2^l
\end{eqnarray}
where
\begin{subequations}\label{e36}
\begin{eqnarray}\label{e36a}
r=\frac{g^2}{\lambda}
\end{eqnarray}
\begin{eqnarray}\label{e36b}
z=\frac{2 m^2}{\lambda\phi^2}
\end{eqnarray}
\begin{eqnarray}\label{e36c}
\Lambda_1=\log\left(\frac{m^2+\lambda\phi^2/2}{\mu^2}\right)
\end{eqnarray}
\begin{eqnarray}\label{e36d}
\Lambda_2=\log\left(\frac{g^2\phi^2}{\mu^2}\right).
\end{eqnarray}
\end{subequations}
The form of Eq. \eqref{e35} follows from several observations.  First of all, $V$ has dimension $[\rm{mass}]^4$, which accounts for the overall factor of $\phi^4$ and the dependence of $p_{L,\ell}^{N,n}$ on $z$.  Secondly, at $L$ loop order, there are $L$ powers of logarithms, either $\Lambda_1$ and $\Lambda_2$, with these two logarithms arising because of the poles in the propagator for the spinor $(g^2\phi^2)$ or the scalar $(m^2 + \lambda\phi^2/2)$.  And finally, at $L$ loop order, there are in general at least $L + 1$ powers of the couplings $\lambda$ and $g^2$, accounting for the overall factor of $\lambda^{N-n} g^{2n}$ and the dependence of $p_{L,\ell}^{N,n}$ on $r$.  

The argument just presented to justify the form of our ansatz in Eq. (35) is that employed in refs. \cite{r37,r38} to justify an expansion of $V$ in massive scalar models.

Following the approach of ref. \cite{r15} we now write
\begin{eqnarray}\label{e37}
\Lambda_2&=&\log\left(\frac{g^2\phi^2}{m^2+\frac{\lambda\phi^2}{2}}\right)+
\log\left(\frac{m^2+\frac{\lambda\phi^2}{2}}{\mu^2}\right) \nonumber \\
&=& \log\left(\frac{2 r}{1+z}\right) + \Lambda_1
\end{eqnarray}
and so Eq. \eqref{e35} can be rewritten as
\begin{eqnarray}
V&=&\phi^4 \sum_{N=L+1}^{\infty} \sum_{L=0}^{\infty} \lambda^N \Pi_L^N(r,z) \Lambda_1^L \label{e38} \\
&=& \phi^4 \sum_{N=1}^{\infty} \lambda^N F_{(N)}(r,z,\xi) \label{e39}
\end{eqnarray}
where $\xi = \lambda \Lambda_1$ and
\begin{eqnarray}\label{e40}
F_{(N)}=\sum_{L=0}^{\infty} \Pi_L^{N+L}(r,z) \xi^L.
\end{eqnarray}
In Eq. \eqref{e39}, $F_{(p)}$ is the ${\rm N}^p\rm{LL}$ contribution to $V$.

Taking the RG equation for this model to be
 \begin{eqnarray}\label{e41}
\left( \mu^2\frac{\partial}{\partial\mu^2}+\beta_\lambda\frac{\partial}{\partial\lambda}+\beta_g\frac{\partial}{\partial g}
+m^2\gamma_m\frac{\partial}{\partial m^2}+\phi^2\gamma_\phi\frac{\partial}{\partial\phi^2}  \right) V=0
\end{eqnarray}
(ignoring the cosmological term \cite{r14,r15,r16,r24,r27})
\begin{eqnarray}\label{e42}
\left[-\lambda \frac{\partial}{\partial \xi} + \beta_\lambda\left(\frac{\partial}{\partial\lambda}
+\left(\frac{\xi}{\lambda}+\frac{1}{z+1}\right) \frac{\partial}{\partial\xi}-\frac{r}{\lambda} \frac{\partial}{\partial r}
-\frac{z}{\lambda} \frac{\partial}{\partial z}\right) \right. \nonumber \\
+\beta_g\left(\frac{2r}{g} \frac{\partial}{\partial r}\right) + \gamma_m\left(z \frac{\partial}{\partial z}+
\frac{\lambda z}{z+1} \frac{\partial}{\partial\xi}\right) \nonumber \\
\left. + \gamma_\phi\left(2-z \frac{\partial}{\partial z}+\frac{\lambda}{z+1} \frac{\partial}{\partial\xi}\right)\right]
\left[\phi^4\sum_{N=1}^{\infty} \lambda^N F_{(N)}(r,z,\xi)\right]=0
\end{eqnarray}
From Ref. \cite{r24} we have the one-loop results
\addtocounter{equation}{1} 
\begin{equation}
\kappa\gamma_m = \frac{\lambda}{2}(1 + 4r),\quad \kappa\beta_\lambda = \frac{1}{2} \lambda^2 (1 + 8r - 48r^2), \quad \kappa\gamma_\phi = -2\lambda r, \quad \kappa\beta_g = \frac{5}{2} g\lambda r;\;\; (\kappa = 16\pi^2) \tag{\theequation a--d} \label{e43}  
\end{equation}
Using Eqs. \eqref{e43} we find that if Eq. \eqref{e42} is satisfied to order $\lambda^2$ we obtain the following for the LL sum $F_{(1)}$
\begin{eqnarray}\label{e44}
\left[\left(\frac{1-48 r^2}{2\kappa}\right)+\left(-1+\frac{1+8 r-48r^2}{2\kappa}\xi\right)\frac{\partial}{\partial\xi} -\frac{1}{2\kappa}\left(1-2r-48r^2\right)r\frac{\partial}{\partial r}\right. \nonumber \\
+\left.\frac{z}{2\kappa}\left(48r^2\right)\frac{\partial}{\partial z}\right] F_{(1)} = 0
\end{eqnarray}
with the boundary condition
\begin{eqnarray}\label{e45}
\lambda F_{(1)}(r,z,\xi=0) \phi^4 &=& \lambda \Pi_0(r,z) \phi^4 \nonumber \\ 
&=& \frac{m^2\phi^2}{2} + \frac{\lambda\phi^4}{4!} \nonumber \\
&=& \lambda\left(\frac z 4 + \frac{1}{24}\right)\phi^4
\end{eqnarray}
which is just the tree level potential. The boundary condition for $F_{(2)}$ would come from the one-loop correction to $V$.

One can now expand $F_{(1)}$ itself so that  
\begin{eqnarray}\label{e46}
F_{(1)}(r,z,\xi)=\sum_{L=0}^{\infty} \Pi_L^{L+1}(r,z) \xi^L ;
\end{eqnarray}
Eqs. (\ref{e44},\ref{e46}) together show that
\begin{eqnarray}\label{e47}
\Pi_{L+1}^{L+2}=\frac{1}{L+1}\left[L \alpha(r,z)+\beta(r,z)+\gamma_r(r,z)\frac{\partial}{\partial r}+\gamma_z(r,z)\frac{\partial}{\partial z} \right]\Pi_L^{L+1}
\end{eqnarray}
where
\begin{equation}
\alpha = \frac{1-r+48r^2}{2\kappa}, \quad \beta = \frac{1-48r^2}{2\kappa}, \quad
\gamma_r = \frac{r\left(-\frac{1}{2} + r+24r^2\right)}{\kappa}, \quad
\gamma_z = \frac{24r^2z}{\kappa}.    \addtocounter{equation}{1} \tag{\theequation a--d} \label{e48}  
\end{equation}
Upon defining $\bar r(t)$, $\bar z(t)$ so that
\begin{equation}
\frac{d\bar{r}}{dt} = \gamma_r(\bar{r}, \bar z), \qquad 
 \frac{d\bar{z}}{dt} = \gamma_z(\bar{r}, \bar z)    \addtocounter{equation}{1} \tag{\theequation a--b} \label{e49}  
\end{equation}
with $\bar r(0) = r$ , $\bar t(0) =0$, and
\begin{eqnarray}\label{e50}
\Phi_L^{L+1}(\bar r, \bar z)=\left[\exp\int_0^t d\tau \left(\alpha(\bar r,\bar z) L + \beta(\bar r,\bar z)\right)\right]\Pi_L^{L+1}(\bar r,\bar z) 
\end{eqnarray}
then by Eqs. \eqref{e48} and \eqref{e50}
\begin{eqnarray}\label{e51}
\Phi_{L+1}^{L+2}\left(\bar r(t(U)),\bar z(t(U))\right) &=& \frac{1}{L+1} \frac{d}{dU} \Phi_L^{L+1} \nonumber \\
&=& \frac{1}{(L+1)!} \frac{d^{L+1}}{dU^{L+1}}\Phi_0^1 
\end{eqnarray}
where
\begin{eqnarray}\label{e52}
U = \int_0^{t(U)} d\tau \exp\left[-\int_0^{\tau} d\sigma(\alpha(\bar r(\sigma),\bar z(\sigma))\right].
\end{eqnarray}
Upon combining Eqs. \eqref{e46}, \eqref{e47}, \eqref{e50} and \eqref{e51} we find that
\begin{eqnarray}\label{e53}
V_{LL}&=&\lambda\sum_{L=0}^{\infty} \Pi_{L}^{L+1}\left(\bar r(t),\bar z(t)\right) \xi^L\phi^4 \nonumber \\
&=& \lambda\phi^4\sum_{L=0}^{\infty} \exp\left(-\int_0^t d\tau \beta\right)\left(\xi\exp-\int_0^td\tau \alpha\right)^L 
\frac{1}{L!}\frac{d^L}{dU^L} \Phi_0^1\left(\bar r(t(U)), \bar z(t(U))\right) \nonumber \\
&=& \lambda\phi^4\exp\left(-\int_0^td\tau\beta\right)\Phi_0^1\left(\bar r(t(U+\Omega)),\bar z(t(U+\Omega))\right)
\end{eqnarray}
where
\begin{eqnarray}\label{e54}
\Omega = \xi\exp\left(-\int_0^t d\tau\alpha\right).
\end{eqnarray}
Upon setting $t=0$, Eq. \eqref{e53} reduces to
\begin{eqnarray}
V_{LL} &=& \lambda \Phi_0^1 (\overline{r}(t(\xi)), 
\overline{z}(t(\xi))\phi^4\nonumber 
\end{eqnarray}
or, by Eq. \eqref{e45}
\begin{eqnarray}\label{e55}
\;\;\;\;\;&=& \lambda \left( \frac{\overline{z}(t(\xi))}{4} + \frac{1}{24} \right) \phi^4.
\end{eqnarray}
To obtain $\bar z(t)$ we return to Eqs. (48c,d; 49a,b);   
\begin{eqnarray}
\frac{d \bar r}{dt}=\frac{\bar r}{\kappa}\left(-\frac 1 2+\bar r+ 24\bar r^2\right) \nonumber
\end{eqnarray}
integrates to give
\begin{eqnarray}\label{e56}
t = \frac{2\kappa}{7}\left[\log\left(\frac{\left(\bar r-\frac 1 8\right)^4\left(\bar r+\frac 1 6\right)^3}{\bar r^7} \frac{r^7}{\left(r-\frac 1 8\right)^4\left(r+\frac 1 6\right)^3}\right)\right]
\end{eqnarray}
and now as
\begin{eqnarray}\label{e57}
\frac{d\bar r}{d\bar z} = \frac{-\frac 1 2 + \bar r+ 24\bar r^2}{24\bar r^2 \bar z}
\end{eqnarray}
we find
\begin{eqnarray}\label{e58}
\left(\frac{\bar z}{z}\right)^7=\left(\frac{\bar r+\frac 1 6}{r+\frac 1 6}\right)^4\left(\frac{\bar r-\frac 1 8}{r-\frac 1 8}\right)^3.
\end{eqnarray}
From Eqs. (\ref{e56}, \ref{e58}) we can implicitly obtain
\begin{eqnarray}
\bar z(\xi) =\bar z\left(t\left(\lambda\log\frac{m^2+\frac{\lambda\phi^2}{2}}{\mu^2}\right) \right)\nonumber
\end{eqnarray}
in \eqref{e55}; that is, we have $V_{LL}$.(The function $t(U)$ is defined in Eq. \eqref{e52}.)

We now will re-express the sum in Eq. \eqref{e38} as
\begin{eqnarray}\label{e59}
V=\phi^4\sum_{L=0}^{\infty} A_L(\lambda,r,z)\Lambda_1^L
\end{eqnarray}
in much the same way as Eq. \eqref{e15} was introduced in the massless self-interacting scalar model. Substitution of Eq. \eqref{e59} into Eq. \eqref{e41} result in
\begin{eqnarray}\label{e60}
&\displaystyle{
\left[-1+\left(\frac{\beta_\lambda}{\lambda}\gamma_\phi+z\gamma_m\right)\frac{1}{1+z}\right](L+1)A_{L+1}}\nonumber \\
&\displaystyle{
+\left[\beta_\lambda\left(\frac{\partial}{\partial\lambda}-\frac{r}{\lambda}\frac{\partial}{\partial r} - \frac{z}{\lambda}\frac{\partial}{\partial z}\right)+\beta_g\left(\frac{2r}{g}\frac{\partial}{\partial r}\right)+\gamma_{\phi}\left(2-z\frac{\partial}{\partial z}\right)+\gamma_m\left(z\frac{\partial}{\partial z}\right)\right]A_L=0}.
\end{eqnarray}
Just as Eq. \eqref{e21} leads to Eq. \eqref{e22}, we see that
\begin{eqnarray}\label{e61}
\left[\left(2-z\frac{\partial}{\partial z}\right)A_L+\frac{L+1}{1+z}A_{L+1}\right] v^2=0,
\end{eqnarray}
and so Eq. \eqref{e21} either implies that $v=0$ (so that the vev of $\phi$ is zero) or
\begin{eqnarray}\label{e62}
B_{L+1}=\frac{1}{L+1} \frac{\partial}{\partial \zeta} B_L\equiv\frac{1}{(L+1)!} \frac{\partial^{L+1}}{\partial\zeta^{L+1}} B_0
\end{eqnarray}
where
\begin{subequations}\label{e63}
\begin{eqnarray}\label{e63a}
\zeta=\log\left(\frac{z}{z+1}\right)\;\;\;(z={\rm e}^\zeta/(1-{\rm e}^\zeta))
\end{eqnarray}
\begin{eqnarray}\label{e63b}
B_L=\exp\left(-2\int^\zeta\frac{d\zeta^\prime}{1-{\rm e}^{\zeta^\prime}}\right)A_L.
\end{eqnarray}
\end{subequations}
Together, Eqs. (\ref{e59},\ref{e62},\ref{e63}) result in
\begin{eqnarray}\label{e64}
V = \phi^4\sum_{L=0}^{\infty} \left(\exp 2\int^\zeta \frac{d\zeta^\prime}{1-{\rm e}^{\zeta^\prime}}\right)\left(\frac{1}{L!}\frac{\partial}{\partial\zeta^L}\right)\left[\exp\left(-2\int^\zeta \frac{d\zeta^\prime}{1-{\rm e}^{\zeta^\prime}}\right)A_0(\lambda,r,z(\zeta))\right]\Lambda_1^L
\end{eqnarray}
but by Eq. \eqref{e63a}
\begin{subequations}\label{e65}
\begin{eqnarray}\label{e65a}
\zeta+\Lambda_1 = \log\left(\frac{z}{1+z}\right) + \log\left(\frac{m^2+\lambda\frac{\phi^2}{2}}{\mu^2}\right)=\log\left(\frac{m^2}{\mu^2}\right)
\end{eqnarray}
and
\begin{eqnarray}\label{e65b}
\exp -2\int_\zeta^{\zeta+\Lambda_1} \frac{d\zeta^\prime}{1-{\rm e}^{\zeta^\prime}}=\left(\frac{\lambda\phi^2}{2(\mu^2-m^2)}\right)^{-2}
\end{eqnarray}
\end{subequations}
and so, by Eq. \eqref{e65}, Eq. \eqref{e64} becomes
\begin{eqnarray}\label{e66}
V = \left(\frac{2(\mu^2-m^2)}{\lambda}\right)^2 A_0\left(\lambda,g,z\left(\log\frac{m^2}{\mu^2}\right)\right).
\end{eqnarray}
As with Eq. \eqref{e33}, $V$ in Eq. \eqref{e66} is independent of $\phi$ -- 
the potential is again ``flat'' with the vev not being determined by the location of a local minimum of $V$ if it is non-zero.

As a check on the validity of Eq. \eqref{e66}, we note that Eqs. \eqref{e60} and \eqref{e61} imply that
\begin{subequations}
\begin{eqnarray}\label{e67a}
A_1 = \frac{1}{\left[1-\left(\frac{\beta_\lambda}{\lambda}+\gamma_\phi+z\gamma_m\right)\frac{1}{z+1}\right]}
\left[\beta_\lambda\left(\frac{\partial}{\partial\lambda}-\frac{r}{\lambda}\frac{\partial}{\partial r}-\frac{z}{\lambda}\frac{\partial}{\partial z}\right)\frac{2r}{g}\beta_g\frac{\partial}{\partial r}\right. \nonumber \\
\left.+z\gamma_m\frac{\partial}{\partial z} + \gamma_{\phi}\left(2-z\frac{\partial}{\partial z}\right)\right]A_0
\end{eqnarray}
and
\begin{eqnarray}\label{e67b}
A_1=(z+1)\left(z\frac{\partial}{\partial z}-2\right)A_0
\end{eqnarray}
\end{subequations}
By combining Eqs. \eqref{e67a} and \eqref{e67b} we arrive at a differential equation for $A_0(\lambda,r,z)$; upon using this equation and (from Eq. \eqref{e63a})
\begin{eqnarray}\label{e68}
z\left(\log\frac{m^2}{\mu}\right) =   \frac{m^2}{\mu^2-m^2} 
\end{eqnarray}
it follows that $V$ as given by Eq. \eqref{e66} satisfies the RG equation of Eq. \eqref{e41}.

We can employ the techniques used to examine the above Yukawa model to investigate the effective potential in the Standard Model. If we have a single $SU(2)$ Higgs doublet with a classical potential
\begin{eqnarray}\label{e69}
V_{cl} = \frac{m^2}{2} \phi^2 + \frac{\lambda}{24} \phi^4
\end{eqnarray}
and $SU(2)$ coupling $g$, U(1) coupling $g^\prime$, $SU(3)$ coupling $g_3$ and top quark Yukawa coupling $h$, then the following logarithms appear in radiative corrections to the effective potential
\begin{equation}
\begin{array}{ll}
\Lambda_1 = \log \left( \displaystyle{\frac{m^2+\lambda\phi^2/2}{\mu^2}}\right),\quad 
\Lambda_2 = \log \left( \displaystyle{\frac{m^2+\lambda\phi^2/6}{\mu^2}}\right),  &
\nonumber \\ & \nonumber \\ 
\Lambda_3 = \log \left( \displaystyle{\frac{g^2\phi^2}{\mu^2}}\right), \quad
\Lambda_4 = \log \left( \displaystyle{\frac{(g^2+g^{\prime^2})\phi^2}{\mu^2}}\right), \quad 
\Lambda_5 = \log \left( \displaystyle{\frac{h^2\phi^2}{\mu^2}}\right). &
\end{array} \addtocounter{equation}{1} \tag{\theequation a--e} \label{e70}    
\end{equation}
Just as was done in the massive Yukawa model in Eq. \eqref{e37}, we can reduce all of these logarithms to $\Lambda_1$. For example, we can write
\begin{eqnarray}\label{e71}
\Lambda_5 &=& \log\left(\frac{h^2\phi^2}{m^2+\displaystyle{\frac{\lambda\phi^2}{2}}}\right) + \Lambda_1 \nonumber \\
&=& \log\left(\frac{2 \rho_h}{1+z}\right) + \Lambda_1
\end{eqnarray}
where
\begin{equation}
\rho = \frac{g^2}{\lambda}, \quad  \rho^\prime = \frac{g^{\prime^2}}{\lambda}, \quad
 \rho_3 = \frac{g^2_3}{\lambda}, \quad \rho_h = \frac{h^2}{\lambda},\quad 
 z = \frac{2 m^2}{\lambda\phi^2}. \addtocounter{equation}{1} \tag{\theequation a--e} \label{e72}     
\end{equation}
The analogue of Eq. \eqref{e38} now becomes
\begin{eqnarray}\label{e73}
V=\phi^4\sum_{N=L+1}^{\infty} \sum_{L=0}^{\infty} \lambda^N\Pi_L^{N}\left(\rho,\rho^\prime,\rho_3,\rho_h,z\right)\Lambda_1^L;
\end{eqnarray}
the LL contribution to V is
\begin{eqnarray}\label{e74}
V_{LL}=\phi^4\sum_{N=0}^{\infty} \lambda^{N+1}\Pi_N^{N+1}\left(\rho,\rho^\prime,\rho_3,\rho_h,z\right)\Lambda_1^N.
\end{eqnarray}
The RG equation
\begin{eqnarray}\label{e75}
\left( \mu \frac{\partial}{\partial\mu}+\beta_\lambda\frac{\partial}{\partial\lambda}+\beta_g\frac{\partial}{\partial g}
+\beta_{g^\prime}\frac{\partial}{\partial g^\prime}+\beta_{g_3}\frac{\partial}{\partial g_ 3}+\beta_h\frac{\partial}{\partial h}
+m^2\gamma_{m^2}\frac{\partial}{\partial m^2}-\phi \gamma_\phi\frac{\partial}{\partial\phi}  \right) V=0
\end{eqnarray}
when used in conjunction with the one-loop RG functions \cite{r28} leads to an equation of the form
\begin{eqnarray}\label{e76}
\Pi_{N}^{N+1}=\frac{1}{N}\left[\alpha N + \beta + r\frac{\partial}{\partial\rho} + r^\prime\frac{\partial}{\partial\rho^\prime} + r_3\frac{\partial}{\partial\rho_3} + r_h\frac{\partial}{\partial\rho_h} + \zeta\frac{\partial}{\partial z}\right]\Pi_{N-1}^N
\end{eqnarray}
provided we keep only those terms pertinent to $V_{LL}$ and discard those that couple $V_{LL}$ to $V_{NLL}$. In Eq. \eqref{e76} we have defined
\begin{subequations}\label{e77}
\begin{eqnarray}\label{e77a}
2\kappa\alpha = 4+12\rho_h-9\rho-3\rho^\prime-36\rho_h^2+\frac 9 4{\rho^\prime}^2+\frac 9 2 \rho\rho^\prime+\frac{27}{4}\rho^2
\end{eqnarray}
\begin{eqnarray}\label{e77b}
2\kappa\beta=-12 \rho_h+9 \rho+3\rho^\prime
\end{eqnarray}
\begin{eqnarray}\label{e77c}
2\kappa r= -\rho\left(4+12 \rho_h-\frac 8 3\rho-3\rho^\prime-36 {\rho_h}^2+\frac 9 4{\rho^\prime}^2+\frac 9 2 \rho\rho^\prime+\frac{27}{4}\rho^2\right) 
\end{eqnarray}
\begin{eqnarray}\label{e77d}
2\kappa r^\prime= -\rho^\prime\left(4+12 \rho_h-9\rho-\frac{50}{3}\rho^\prime-36 {\rho_h}^2+\frac 9 4{\rho^\prime}^2+\frac 9 2 \rho\rho^\prime+\frac{27}{4}\rho^2\right) 
\end{eqnarray}
\begin{eqnarray}\label{e77e}
2\kappa r_3= -\rho_3\left(4+12 \rho_h-9\rho-{3}\rho^\prime+14\rho_3-36 {\rho_h}^2+\frac 9 4{\rho^\prime}^2+\frac 9 2 \rho\rho^\prime+\frac{27}{4}\rho^2\right) 
\end{eqnarray}
\begin{eqnarray}\label{e77f}
2\kappa r_h=-\rho_h\left(4+3 \rho_h-\frac 9 2\rho-\frac{1}{6}\rho^\prime+16\rho_3-36 {\rho_h}^2+\frac 9 4{\rho^\prime}^2+\frac 9 2 \rho\rho^\prime+\frac{27}{4}\rho^2\right) 
\end{eqnarray}
\begin{eqnarray}\label{e77g}
2\kappa \zeta =-z\left(2-36 {\rho_h}^2+\frac 9 4{\rho^\prime}^2+\frac 9 2 \rho\rho^\prime+\frac{27}{4}\rho^2\right). 
\end{eqnarray}
\end{subequations}
Just as Eq. \eqref{e47} leads to Eq. \eqref{e55}, Eq. \eqref{e76} leads to
\begin{eqnarray}\label{e78}
V_{LL}=\frac{\lambda}{4}\left(\bar z(t(\lambda \Lambda_1))+\frac 1 6\right)\phi^4
\end{eqnarray}
where (as in Eq. \eqref{e48})
\begin{equation}
\frac{d\bar{\rho}(t)}{dt} = \bar{r}\left( \bar{\rho}(t), \bar{\rho}^\prime(t), \bar{\rho}_3(t), \bar{\rho}_h(t), \bar{z}(t)\right) \addtocounter{equation}{1} \tag{\theequation a} \label{e79a}   
\end{equation}
\[ \hspace{-3cm}\vdots \]
\begin{equation}
\frac{d\bar{z}(t)}{dt} = \bar{\zeta}\left( \bar{\rho}(t), \bar{\rho}^\prime(t), \bar{\rho}_3(t), \bar{\rho}_h(t), \bar{z}(t)\right)  \tag{\theequation e} \label{e79e}   
\end{equation}
%
with $\bar\rho(0) = \rho$ etc, and
\begin{eqnarray}\label{e80}
U=\int_0^{t(U)} d\tau \exp\left[\int^\tau d\sigma\left(\alpha(\bar\rho(\sigma),\dots ,\bar z(\sigma))\right)\right].
\end{eqnarray}
so that if $t=0$, $U=0$ also. Exact determination of $\bar z(t(\lambda\Lambda_1))$is prohibitively difficult, but if we assume that $\lambda$ is the only coupling of consequence, then by Eq. \eqref{e77g} 
\begin{eqnarray}\label{e81}
\frac{\bar z(t)}{dt}=-\frac{\bar z(t)}{\kappa} \implies \bar z(t)=\frac{2 m^2}{\lambda\phi^2}\displaystyle{{\rm e}^{-\frac{t}{\kappa}}}
\end{eqnarray}
and by Eqs. \eqref{e77a} and \eqref{e80}
\begin{eqnarray}\label{e82}
U=\int_0^{t(U)} d\tau \exp\left(\int_0^\tau d\sigma \frac{2}{\kappa}\right) \implies U=\frac{\kappa}{2}\left({\rm e}^{2t(U)}{\kappa} -1 \right). 
\end{eqnarray}
Eqs. \eqref{e81} and \eqref{e82} lead to
\begin{eqnarray}\label{e83}
\bar z(t(\lambda\Lambda_1))=\frac{2 m^2}{\lambda \phi^2}\left(1+\frac{2}{\kappa}\lambda\Lambda_1\right)^{-1/2}
\end{eqnarray}
and so Eq. \eqref{e78} becomes
\begin{eqnarray}\label{e84}
V_{LL}=\frac{\lambda\phi^4}{24}+\frac{m^2\phi^2}{2}\left[1+\frac{2}{\kappa}\lambda\log\left(\frac{m^2+ \lambda\phi^2}{\mu^2}\right)\right]^{-\frac{1}{2}}.
\end{eqnarray}

The term in the derivative expansion of the effective action beyond the effective potential is $\frac 1 2 Z(\phi)(\partial_\mu\phi)^2$ \cite{r1} with $Z$ at classical level being
\begin{eqnarray}\label{e85}
Z_{cl}=1.
\end{eqnarray}
One can apply the arguments applied to $V$ above to compute the LL contribution to $Z$ in the Standard Model. In much the same way that Eq. \eqref{e78} follows from Eq. \eqref{e69},  it follows that Eq. \eqref{e85} leads to
\begin{eqnarray}\label{e86}
Z_{LL} = 1.
\end{eqnarray}
This is consistent with the result given in ref \cite{r29}.

We can also expand the effective potential in the Standard Model in the same way Eq. \eqref{e59} was used to expand the effective potential in the massive Yukawa model
\begin{eqnarray}\label{e87}
V=\phi^4\sum_{L=0}^{\infty} A_L(\lambda,\rho,\rho^\prime,\rho_3,\rho_h,z) \Lambda_1^L.
\end{eqnarray}
Again applying Eq. \eqref{e21} we are led to Eq. \eqref{e61}; we then are led to
\begin{eqnarray}\label{e88}
V = \left[\frac{2(\mu^2-m^2)}{\lambda}\right]^2 A_0\left(\lambda,\rho,\rho^\prime,\rho_3,\rho_h,z\left(\log\frac{m^2}{\mu^2}\right)\right)
\end{eqnarray}
in the Standard Model, just as Eq. \eqref{e66} arises in  a massive Yukawa model. Once again, $V$ has no dependence on $\phi$ provided the vev of $\phi$ is non-zero, and this vev is not determined by a local minimum of the potential.

\section{Discussion}

We have discussed the effective potential in a massless self-interacting scalar model, in a Yukawa model and the non-conformal Standard Model. In all three cases, the RG equation can be used to sum ${\rm N}^p {\rm LL}$ contributions to the effective potential. In addition by applying the minimization condition of Eq. \eqref{e21} it follows that either the vev of $\phi$ vanishes, or the effective potential is independent of $\phi$.  It is possible that a second set of scalars or a Stueckelberg mass for a $U(1)$ gauge field \cite{r7} could alter this situation. It does appear though that if there is a single scalar field coupled to other fields (spinor or vector), radiative effects lead to either a flat potential or no spontaneous symmetry breaking, irrespective of what happens classically.

In the latter two models we have several coupling constants and several mass scales. We have followed  Ref. \cite{r15,r16} and re-expressed all logarithms in terms of $\Lambda_1$ and used a single RG mass scale $\mu$; in principle a logarithm other than $\Lambda_1$ could have been employed. In the Standard Model we have taken $\lambda$ to be the dominant coupling, larger than $g$,  $g^\prime$, $g_3$ or $h$. This is consistent with the results of Refs. \cite{r9,r10}.
We note that in Ref \cite{r1}, Eq. \eqref{e61} is used in conjunction with the one loop contribution to $V$ in massless scalar electrodynamics to fix a relationship between the gauge and quartic scalar coupling at mass scale $\mu=v$; we have used it to relate $A_{M+1}$ to $A_M$ without relating the couplings (which, after all, should be independent).

There is an extensive discussion in the literature of the implications of refs. \cite{r32,r33} which state that the exact effective potential should be both real and convex. (A discussion of this also occurs in ref. \cite{r34}.)  
By ``convex'' we mean that 
\begin{equation}\label{e89}
V(x\phi_1 + (1-x)\phi_2) \le x V(\phi_1)  + (1-x) V(\phi_2) 
\end{equation}
for  $0\le x \le 1$ for all $\phi_1$ and $\phi_2$. Consequently, $V$ should satisfy the condition that a linear interpolation of $V(\phi)$
is always larger than or equal to $V(\phi)$. 
The one loop result of \eqref{e34} will not be consistent with this result if $m^2 < 0$ and also $-\frac{\lambda\phi^2}{2} > m^2$.  In this case $V^{(1)}$ is no longer convex and develops an imaginary part.  A paper that deals with this problem is \cite{r35}; there it is assumed that in the region between two minima of the perturbative effective potential $V$  where $V^{\prime\prime} < 0$, the true effective potential becomes ``flat'' in time, thereby becoming convex and stable.  The same conclusion is met in \cite{r36}.  
Other aspects of this problem are considered in ref. \cite{r41}.
We note that in refs. \cite{r39,r40} a ``Maxwell construction'' is employed to ensure that $V$ is convex in the region between two minima; its contribution is attributed 
to non-perturbative effects
that arise as a result of summation over all saddle point contributions in the path integral computation of the effective potential. 
In the limit of infinite volume the potential becomes flat in this region.  The exact results we have arrived at above (Eqs. \eqref{e31}, \eqref{e66}, \eqref{e88}) 
in which $V(\phi)$ is ``flat'' (ie, $V^\prime (\phi) = V^{\prime\prime}(\phi) = 0$ for all $\phi$) is automatically consistent with the requirement that $V$ be real and convex for all $\phi$ no matter what $m^2$ is.  Furthermore, we have worked directly from the ansatz of Eq. \eqref{e35} which follows from the form that $V$ takes perturbatively without any additional assumption.  This results in $V$ being analytic (since it is flat) without the discontinuities occurring in the form of $V$ appearing in refs. \cite{r39,r40}.
Our results stem from using the ansatz of Eqs. \eqref{e2} and \eqref{e35} for the general form of $V$. These ansatz are a consequence of the perturbative expansion for $V$
(given, for example, by Eq. (7.6) of ref. \cite{r34}); no additional input coming from non perturbative effects that might lead to non-analytic behaviour for $V$ are considered.
Having the classical ``Mexican hat'' potential (such as occurs in the potential $V = \frac 1 2 m^2 \phi^2 + \frac{1}{4!} \lambda \phi^4$ if $m^2 < 0$ and $\lambda > 0$)
became ``flat'' (once all logarithmic perturbative effects are summed using the RG equation and the minimization condition of Eq. \eqref{e21} is applied) is actually a 
straightforward way to achieving consistency with the general convexity condition of Eq. \eqref{e89}.

As noted in the introduction, having an effective potential that is independent of $\phi$ does not preclude the possibility of interactions involving $\phi$. It is possible that the theory is ``trivial'' and that interactions are only possible in an effective theory which involves a cutoff, or it may possibly be that interactions involve 
only derivatives of $\phi$ or interactions between $\phi$ and other fields.  Answering these questions is a non-trivial problem. 

We have ignored the issue of gauge dependence in our consideration of the effective potential. For a recent discussion of this problem, see Ref. \cite{r30}.

A discussion of the phenomenological consequences of these formal results will be forthcoming. The possible implications of having a flat potential on cosmological models is also being considered.

\section*{Acknowledgements} 
{F. T. B.} would like to thank CNPq for financial support.
\hbox{D. G. C. M.}  would like to thank Roger Macleod for several pertinent observations
and the  {\it Universidade de S\~ao Paulo}  for the hospitality during the realization of this work.
His visit was supported by a FAPESP grant.
 
%

\begin{thebibliography}{10}

\bibitem{r1} 
S.~R. Coleman and E. Weinberg, Phys. Rev. {\bf D7},  1888  (1973).

\bibitem{r2} 
S. Weinberg, Phys. Rev. {\bf D7},  2887  (1973).

\bibitem{r3} 
R. Jackiw, Phys. Rev. {\bf D9},  1686  (1974).

\bibitem{r4} 
A. Salam and J. Strathdee, Phys. Rev. {\bf D9},  1129  (1974).

\bibitem{r5} 
S. Weinberg, {\em The Quantum Theory of Fields} (Cambridge University Press,
  Cambridge, England, 1995).

\bibitem{r6} 
S.~V. Kuzmin and D.~G.~C. McKeon, Mod. Phys. Lett. {\bf A16},  747  (2001).

\bibitem{r7} 
D.~G.~C. McKeon and T.~J. Marshall, Int. J. Mod. Phys. {\bf A23},  741  (2008).

\bibitem{r8} 
V. Elias, R.~B. Mann, D.~G.~C. McKeon, and T.~G. Steele, Phys. Rev. Lett. {\bf  91},  251601  (2003);
%
Nucl. Phys. {\bf  B678},  147  (2004) (Erratum ibid 703, 413 (2004));
%
Phys. Rev. {\bf D72}, 037902  (2005).

\bibitem{r9} 
F.~A. Chishtie, T. Hanif, J. Jia, R.~B. Mann, D.~G.~C. McKeon, T.~N. Sherry,
  and T.~G. Steele, Phys. Rev. {\bf D83},  105009  (2011).

\bibitem{r10} 
T.~G. Steele and Z.-W. Wang, Phys. Rev. Lett. {\bf 110},  151601  (2013).

\bibitem{r11} 
M.~R. Ahmady, F.~A. Chishtie, V. Elias, A.~H. Fariborz, N. Fattahi, D.~G.~C.
  McKeon, T.~N. Sherry, and T.~G. Steele, Phys. Rev. {\bf D66},  014010
  (2002).

\bibitem{r12} 
V. Elias and D.~G.~C. McKeon, Int. J. Mod. Phys. {\bf A18},  2395  (2003).

\bibitem{r13} 
D.~G.~C. McKeon and A. Rebhan, Phys. Rev. {\bf D67},  027701  (2003).

\bibitem{r14} 
B.~M. Kastening, Phys. Lett. {\bf B283},  287  (1992).

\bibitem{r15} 
B.~M. Kastening, hep-ph/9207252  (1992).

\bibitem{r16} 
M. Bando, T. Kugo, N. Maekawa, and H. Nakano, Phys. Lett. {\bf B301},  83
  (1993).

\bibitem{r17} 
S. Weinberg, Phys. Lett. {\bf B82},  387  (1979).

\bibitem{r18} 
L. Susskind, Phys. Rev. {\bf D20},  2619  (1979).

\bibitem{r19} 
F.~T. Brandt, F.~A. Chishtie, and D.~G.~C. McKeon, Mod. Phys. Lett. {\bf A20}, 2215  (2005);
  Int.\ J.\ Mod.\ Phys.\ A {\bf 22} (2007).

\bibitem{r20} 
F.~A. Chishtie, M.~T. Hanif, J. Jia, D.~G.~C. McKeon, and T.~N. Sherry, Int. J.
  Mod. Phys. {\bf A25},  5711  (2010).

\bibitem{r21} 
D. McKeon, Can. J. Phys. {\bf 89},  277  (2011).

\bibitem{r22} 
F.~A. Chishtie, T. Hanif, D.~G.~C. McKeon, and T.~G. Steele, Phys. Rev. D {\bf
  77},  065007  (2008).

\bibitem{r23} 
R. Courant and D. Hilbert, {\em Methods of Mathematical Physics} (Interscience,
  NY, 1966), Vol.~II, Ch. 2.

\bibitem{r24} 
M.~B. Einhorn and D.~T. Jones, Nucl. Phys. {\bf B230},  261  (1984).

\bibitem{r25} 
C. Ford and C. Wiesendanger, Phys. Rev. {\bf D55},  2202  (1997).

\bibitem{r26} 
C. Ford, hep-th/9609127  (1996).

\bibitem{r27} 
B.~M. Kastening, Phys.Rev. {\bf D54},  3965  (1996).

\bibitem{r28} 
C. Ford, D. Jones, P. Stephenson, and M. Einhorn, Nucl.Phys. {\bf B395},  17
  (1993).

\bibitem{r29} 
F. Chishtie, J. Jia, and D. McKeon, Phys.Rev. {\bf D76},  105006  (2007).

\bibitem{r30} 
  A.~Andreassen, W.~Frost and M.~D.~Schwartz,
  Phys.\ Rev.\ {\bf D91},  016009 (2015).


\bibitem{r31} T.G. Steele, Zhi-Wei Wang and D.G.C. McKeon, Phys.Rev. {\bf D90}, 105012 (2014).

\bibitem{r32} K. Symanzik, Comm. Math. Phys. {\bf 16}, 48 (1970).

\bibitem{r33} J. Iliopoulos, C. Itzykson and A. Martin, Rev. Mod. Phys. {\bf 47}, 165 (1975).

\bibitem{r34}
V. Miransky, {\em Dynamical Symmetry Breaking in Quantum Field Theories} (World Scientific, Singapore, 1993) Section 7.7.

\bibitem{r35} M.B. Einhorn and D.R.T. Jones, JHEP {\bf 0704}, 051 (2007).

\bibitem{r36} V. Branchina, P. Castorina and D. Zappala, Phys. Rev. {\bf D41}, 1948 (1990).

\bibitem{r37} B. Kastening, Phys. Lett. {\bf B283}, 287 (1992); hep-phy 9207252.

\bibitem{r38} J.M. Chung and B.K. Chung, Phys. Rev. {\bf D60}, 105001 (1999).

\bibitem{r39} J. Alexandre, Phys. Rev. {\bf D86}, 025030 (2012).

\bibitem{r40} J. Alexandre and A. Tsapalis, Phys. Rev. {\bf D87}, 025028 (2013).

\bibitem{r41} 
 V.~Branchina, H.~Faivre and V.~Pangon,
  J.\ Phys.\ G {\bf 36}, 015006 (2009);
 Y.~Fujimoto, L.~O'Raifeartaigh and G.~Parravicini,
  Nucl.\ Phys.\ B {\bf 212}, 268 (1983);
  R.~W.~Haymaker and J.~Perez-Mercader,
 Phys.\ Rev.\ D {\bf 27} (1983) 1948.
 doi:10.1103/PhysRevD.27.1948

\end{thebibliography}


\end{document}